# A Dataset of Human Motion Status Using IR-UWB Through-wall Radar

**ZHENGLIANG ZHU[1,3], DEGUI YANG[2], JUNCHAO ZHANG[2], and FENG TONG[1,3]**

[1] Key Laboratory of Underwater Acoustic Communication and Marine Information Technology of the Ministry of Education, Xiamen University, Xiamen, Fujian 361005, China.
[2] School of Aeronautics and Astronautics, Central South University, Changsha, Hunan 410083, China.
[3] College of Ocean and Earth Sciences, Xiamen University, Xiamen, Fujian 361005, China.

Corresponding author: DEGUI YANG (e-mail: ah_gui@csu.edu.cn).

This work was supported in part by the National Key Research and Development Program of China under Grant 2018YFC0810202.

**ABSTRACT** Ultra-wideband (UWB) through-wall radar has a wide range of applications in non-contact human information detection and monitoring. With the integration of machine learning technology, its potential prospects include the physiological monitoring of patients in the hospital environment and the daily monitoring at home. Although many target detection methods of UWB through-wall radar based on machine learning have been proposed, there is a lack of an opensource dataset to evaluate the performance of the algorithm. This published dataset was measured by impulse radio UWB (IR-UWB) through-wall radar system. Three test subjects were measured in different environments and several defined motion statuses. Using the presented dataset, we propose a human-motion-status recognition method using a convolutional neural network (CNN), the detailed dataset partition method and recognition process flow is given. On the well-trained network, the recognition accuracy of testing data for three kinds of motion statuses is higher than 99.7%. The dataset presented in this paper considers a simple environment. Therefore, we call on all organizations in the UWB radar field to cooperate to build opensource datasets to further promote the development of UWB through-wall radar.

## I. INTRODUCTION

Ultra-wideband (UWB) through-wall radar is widely used in civil and military fields because of its strong penetrability to non-metal and non-transparent obstacles and high range resolution. IR-UWB through wall radar as a form of UWB through-wall radar, its hardware structure is relatively simple. The transmitted impulse has a wide frequency range and it can overcome the influence of radar blind area. Its applications mainly include: post-disaster rescue [1] (e.g. earthquakes, mudslides, mine disasters, etc.), anti-terrorism rescue [2], [3], non-contact life monitoring [4], [5], etc. As a common non-contact detection technology, radar has advantages in environmental applicability and privacy protection compared with visible light, infrared, and acoustic technologies [6], and has broad application prospects in indoor human monitoring. In recent years, with the improvement of computing power and the development of artificial intelligence technology, machine learning has shown great promises in the application of UWB through-wall radar target recognition.

Aiming at the problem of the wrong refscue caused by common animal breathe frequency range similar to humans in earthquake rescue, the recognition of human and dog based on UWB radar is studied in [7], and a support vector machine (SVM) method based on recursive feature elimination is proposed. The proposed method is designed to optimize the radar echo features of people and dogs to increase the performance of the recognition model. In the through-wall condition, the method proposed in [7] obtains excellent performance to distinguish human and dogs. In [8], the distinguished method between humans and common animals (e.g. cat, dog, rabbit) based on the multiscale residual attention network is proposed. The method relies on the difference in the respiratory signal spectrum between humans and common animals. Therefore, the input data size is analyzed form the perspective of network design and predict performance. If the input data dimension is large, the calculation amount of the distinguished network increases; while the input data dimension is small, it is difficult to reflect the breath information of the human and it will damage the distinguished accuracy. The literature [9]–[11] carry out related work for the recognition of human motion status based on carrier-free UWB radar named SIR-20. A variety of feature extraction methods for radar received signals are proposed, such as 2D-variational mode decomposition (VMD) method, 2D-discrete wavelet transform (DWT) method. In [12], the radar signals of eight human volunteers performing eight different statuses (e.g. walking, running, rotating, punching, jumping, transitioning between standing and sitting, crawling, and standing still) are collected based on UWB through-wall radar. The principal component analysis (PCA) is used to compress the signals and the PCA coefficients adopt to describe the feature of different human statuses. Combined with the SVM classifier, the recognition accuracy of different statuses is higher than 85%.

In [13], the recognition method of human finer-grained activities under the through-wall condition is discussed based on the stepped frequency continuous wave (SFCW) UWB through-wall radar. A comprehensive range accumulation time-frequency transform (CARTFR) based on inverse weight coefficients is proposed to enhance the micro-Doppler features of human finer-grained activity signals. Based on the selected features in the CARTFR spectrum, SVM is adopted to classify different activities. In the experiment with a fixed position behind the wall, the classification accuracy of the 6 activities performed by 8 volunteers is higher than 93.23%. When the human position is changed, the recognition accuracy of the 5 activities with 6 meters behind the wall is 86.67%. The literature [14] proposed a fine-grained micro-Doppler feature extraction method for human activities in through wall environment based on Multiple Hilbert-Huang Transforms (MHHT). Combined with the principle of human kinematics, it is proved that there is a connection between the MHHT features components and human body structure. A Segmented convolutional gated recurrent neural network is adopted to recognize human activity measured by UWB radar in [15]. Unlike some methods used the micro-Doppler spectrograms as input data to train the recognition model, the proposed method extracts the segmented features of micro-Doppler spectrum through convolution operation and uses gate recursive unit to encode the feature maps along the time dimension. The experimental results show that the method can deal with human activities in any length and have better performance in fine temporal resolution, noise robustness, and generalization.

Compared with the computer vision domain (e.g. ImageNet, the dataset includes 1.5 million images [16]), radar human recognition based on machine learning lacks a large radar target dataset. Therefore, related scholars have explored the research of human status recognition methods based on transfer learning [17], [18], small sample learning [19], [20]. In [18], the human status recognition method based on unsupervised domain adaptation. To solve the problem of the insufficient training data in actual measurement, the dataset MOCAP [21] is used to pre-train the recognition network, and via the domain adaptation method combined with measured data to achieve human status recognition. To overcome the problem of unbalanced data, the adaptive incremental recursive least-squares regression parameter estimation method using the UWB radar network is proposed in [19]. The proposed method adaptively segments the dynamic signals and realizes the recognition of the human motion status under a small sample condition via the tensor deep learning model. The accuracy of the recognition of the four-motion status under the condition of three radars networked is 91.62%.

From the previous literature review, we can see that a lot of research work focused on feature extraction, design, and application of machine learning models. However, there is still a problem in the recognition of human status based on impulse radio UWB (IR-UWB) through-wall radar, that is, the lack of public human motion status radar dataset, and it is difficult to compare the performance of existing recognition algorithm on the same baseline. In this paper, a variety of human motion status radar echoes under different penetrating media were collected using the IR-UWB through-wall radar system. Then, the IR-UWB human motion status dataset is constructed after preprocessing. (This dataset has been publicly posted on GitHub, the link will be noted at Supplementary Materials.) Finally, to enable other scholars to conduct related research based on this dataset, we propose a method for recognizing human motion status base on convolutional neural networks (CNN) for the reference.

The paper is organized as follows. In Section II, the human motion model of the UWB radar is briefly introduced and simulated. Based on the feature description of the simulation echoes, the motion statuses of humans in the measured scene are defined. In Section III, the human motion status data in different human bodies and different penetrating media is collected base on IR-UWB through-wall radar system, converted into images and constructed as a dataset. In Section IV, a usage demo is provided based on the dataset using CNN and the specific recognition process is given. Finally, conclusions and prospects are provided in Section V.

## II. THEORY AND MODEL

### A. UWB RADAR HUMAN MOTION MODEL

Since the IR-UWB through-wall radar has a high range resolution, the emitted electromagnetic wave is reflected by the human target and then received by the receiver to form an echo. The human return signal is composed of the superposition of different time delays of human multiple scattering centers (e.g. arm, limb, main torso) and the surrounding environment (e.g. ground, windows, roof). These scattering have different scattering coefficients [22]. Then, the UWB radar received signal can be expressed as follow:

$$r(t) = \sum_{n=1}^{N} A_n p(t - \tau_n) \quad (1)$$

where $p(t)$ denotes the transmit signal of IR-UWB through-wall radar, $r(t)$ denotes the received signal, $N$ is the number of multipath reflection from human scattering centers and the surrounding environment, $A_n$ and $\tau_n$ are the amplitudes and time delay of corresponding scattering center of the human and surrounding environment, respectively. In a short observe time, the received echo signal can be used to characterize the motion statues of humans, the reason is that the different status will cause the parameters in (1) changed. An improved UWB signal model is introduced in [23], in the following we extend the model to simulate UWB human motion received signals. In this simulation, we set the following assumptions: the signal transmitted by the radar does not attenuate during propagation, and the transmitted signal is the Gaussian pulse signal; the transceiver antenna of the radar is in ideal condition; The scattering center of human-only considers the torso and the displacement of chest; the human motion status is walking, standing still.

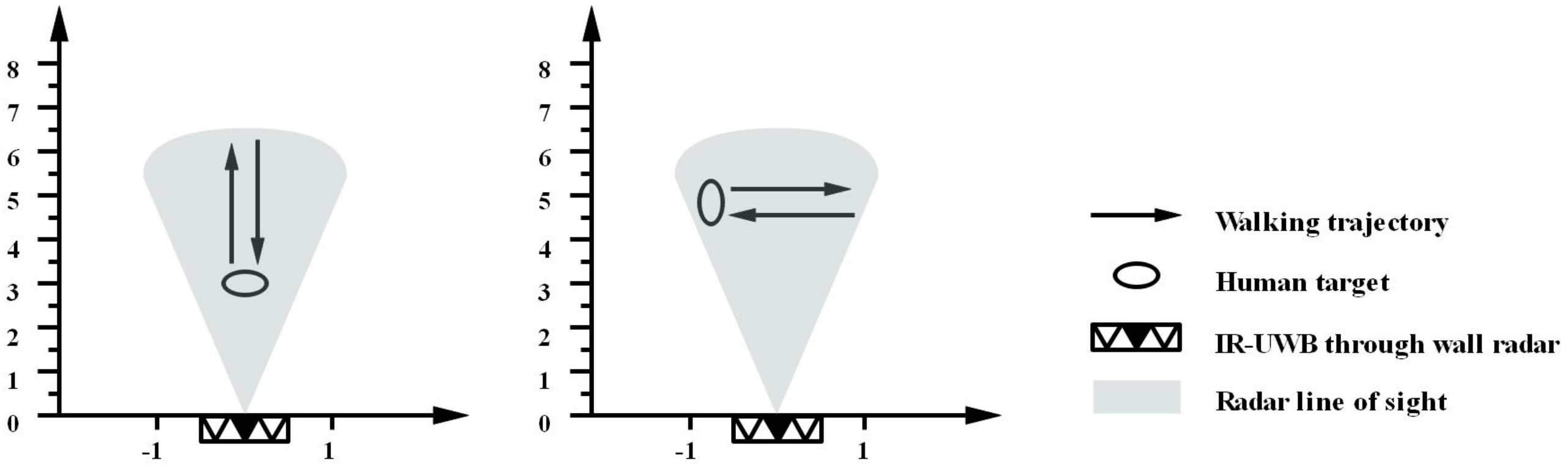

(a)

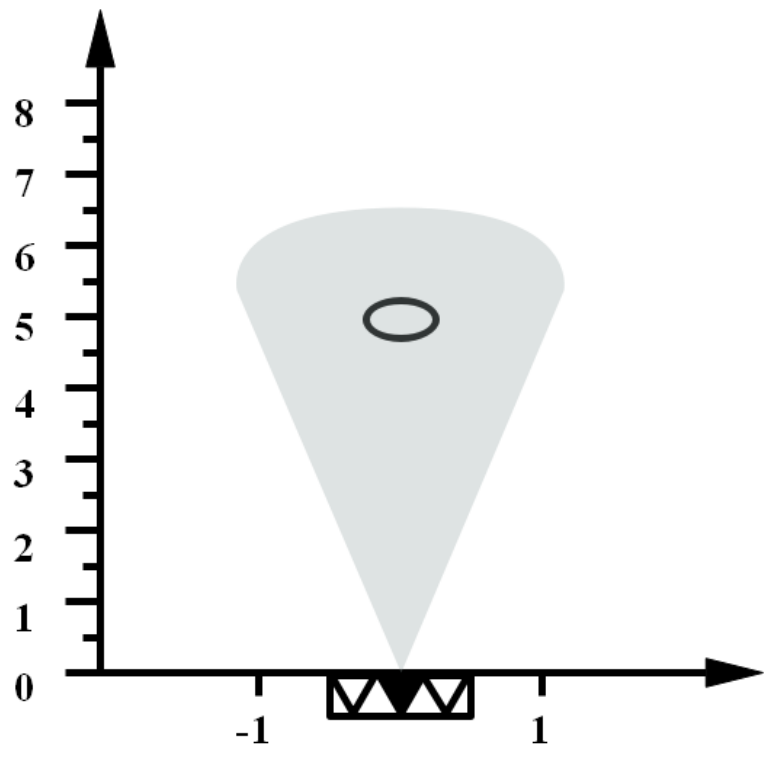

(b)

**FIGURE 1. Schematic diagram of simulation scene. (a) human walking. (b) human standing still.**

The human motion status is shown in FIGURE 1. In (a), considering that the human walking with two trajectories. The first trajectory is walking back and forth along the radar line of sight radially, and its round-trip period is 12 seconds. The second trajectory is walking perpendicular to the radar light of sight tangentially, and its round-trip period is 10 seconds. In (b), considering that the human standing still. The human simulation respiratory frequency is 0.3Hz and corresponding chest placement is 0.05 meters. Then, the simulation receive signals can be expressed:

$$r(t) = p(t - \tau_d(t)) = p(t - \frac{2 \times d(t)}{c}) \quad (2)$$

In the (2), $c = 3 \times 10^8 m/s$ is the propagation speed of electromagnetic waves, $\tau_d(t)$ is the time delay as the human trajectory changes, $d(t)$ is the human trajectory and a function of time $t$. Discrete (2) along the fast time dimension (i.e. range dimension) and slow time dimension, we can obtain the simulation receive signal matrix $R[m,n]$.

$$\begin{aligned} R[m,n] &= r[mT_s, nT_f] \\ &= -A_p \times \frac{2 \times [nT_f - \frac{2 \times d(mT_s)}{c}]}{\rho^2} \times e^{-\frac{[nT_f - \frac{2 \times d(mT_s)}{c}]}{\rho^2}} \end{aligned} \quad (3)$$

where $m$ and $T_s$ represent the sampling points and sampling interval along the slow time direction, respectively; $n$ and $T_f$ represent the sampling points and sampling interval along the fast time direction, respectively; $A_p$ denotes the amplitude of transmit signal, $\rho$ denotes the pulse width parameter of transmit signal.

FIGURE 2 shows the corresponding simulation results. Generally, it is called the time-domain range profile, which characterizes the distance relationship between the human and radar. In other words, the time-domain range profile can be adapted to the basis for recognizing human motion status.

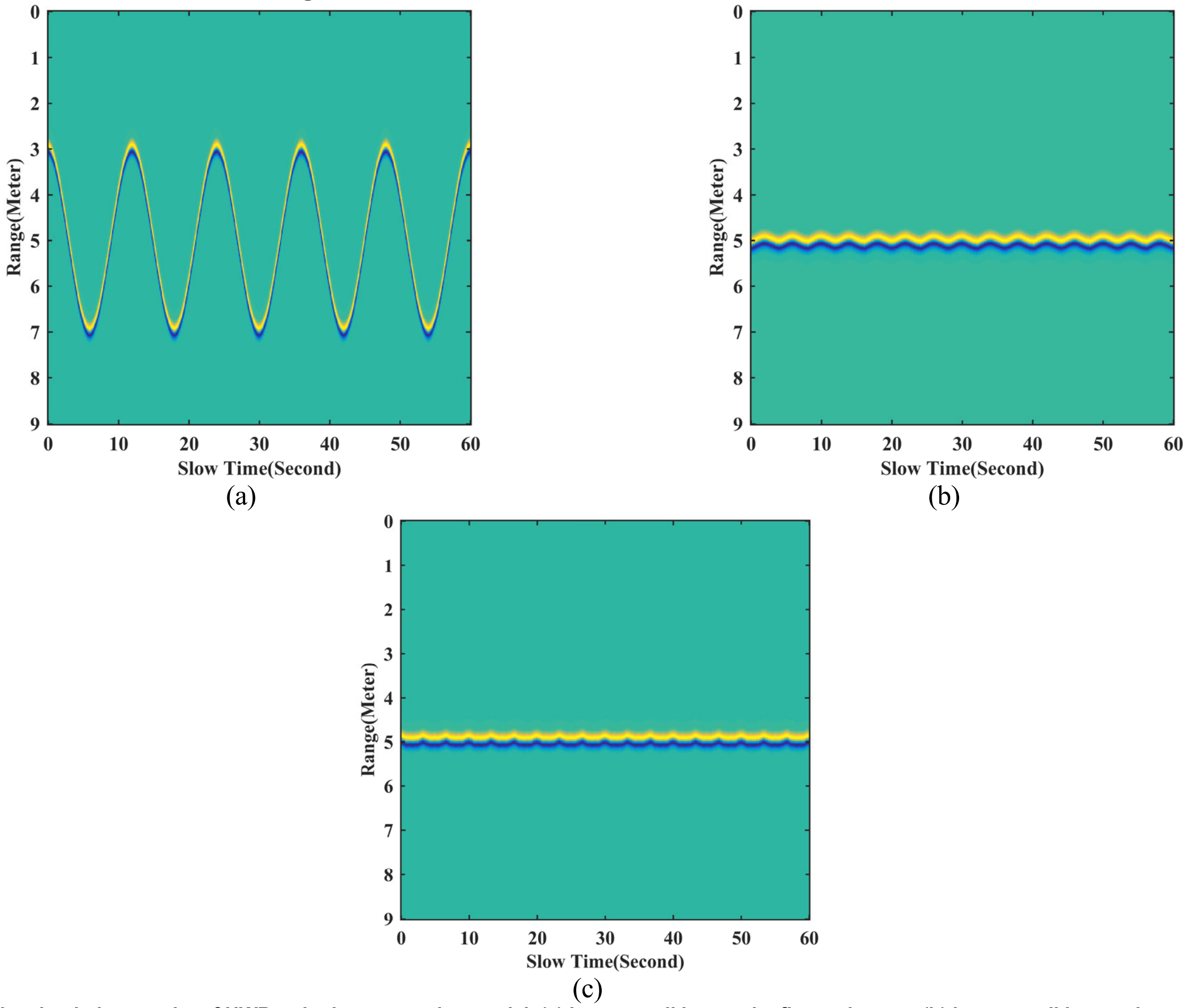

**FIGURE 2.** **The simulation results of UWB radar human motion model. (a) human walking on the first trajectory. (b) human walking on the second trajectory. (c) human standing still.**

### *B. HUMAN TARGET ECHO SIGNALS FEATURE DESCRIPTION AND MOTION STATUS DEFINITION*

According to the research of target motion status recognition based on UWB radar received signals, the different feature description methods can be divided into time-domain range and time-frequency domain based. The main purpose is to analyze the radar echoes characteristics in a short-term observation time window to recognize the human motion status.

Time-frequency domain feature description: analyze radar echoes characteristic via the time-frequency transformation method. (e.g. short-time Fourier transform (STFT), Wigner-Ville distribution (WVD), wavelet transform (WT), Hilbert-Huang transform (HHT)). The paper [24] uses Doppler radar to collect the echoes of human different activity and transforms the signals into the time-frequency domain for feature extraction via STFT. The extracted features and definitions are as follows: 1) torso frequency: reflects the speed of a human, the speed of the human torso varies greatly depending on the human motion status; 2) the total bandwidth (BW) of the Doppler signal: describes the speed of human limb, e.g. arms/legs produce a larger Doppler signal BW; 3) the offset of the total Doppler signal: represents the asymmetry of human limbs between moving forward and backward; 4) the BW without micro-Dopplers: denotes the swing changes of human torso; 5) the period of the limb motion: describes the swing rate of the arm or leg. FIGURE 3 shows the illustration of the features.

Time-domain range profile feature description: take the result of FIGURE 2 as an example, the features extracted are shown in FIGURE 4. The specific features are defined as follows: 1) Period: describes the time needed for the human to complete the walking trajectory. When the human standstill, this value can be approximated as the respiratory cycle; 2) Chest motion range: denotes slight displacement of the chest cavity due to respiration; 3) Human motion range: reflects the max distance between the radar and human; 4) Instantaneous velocity angle: describes the instantaneous velocity of the human, and the greater the human velocity is, the greater the value will be.

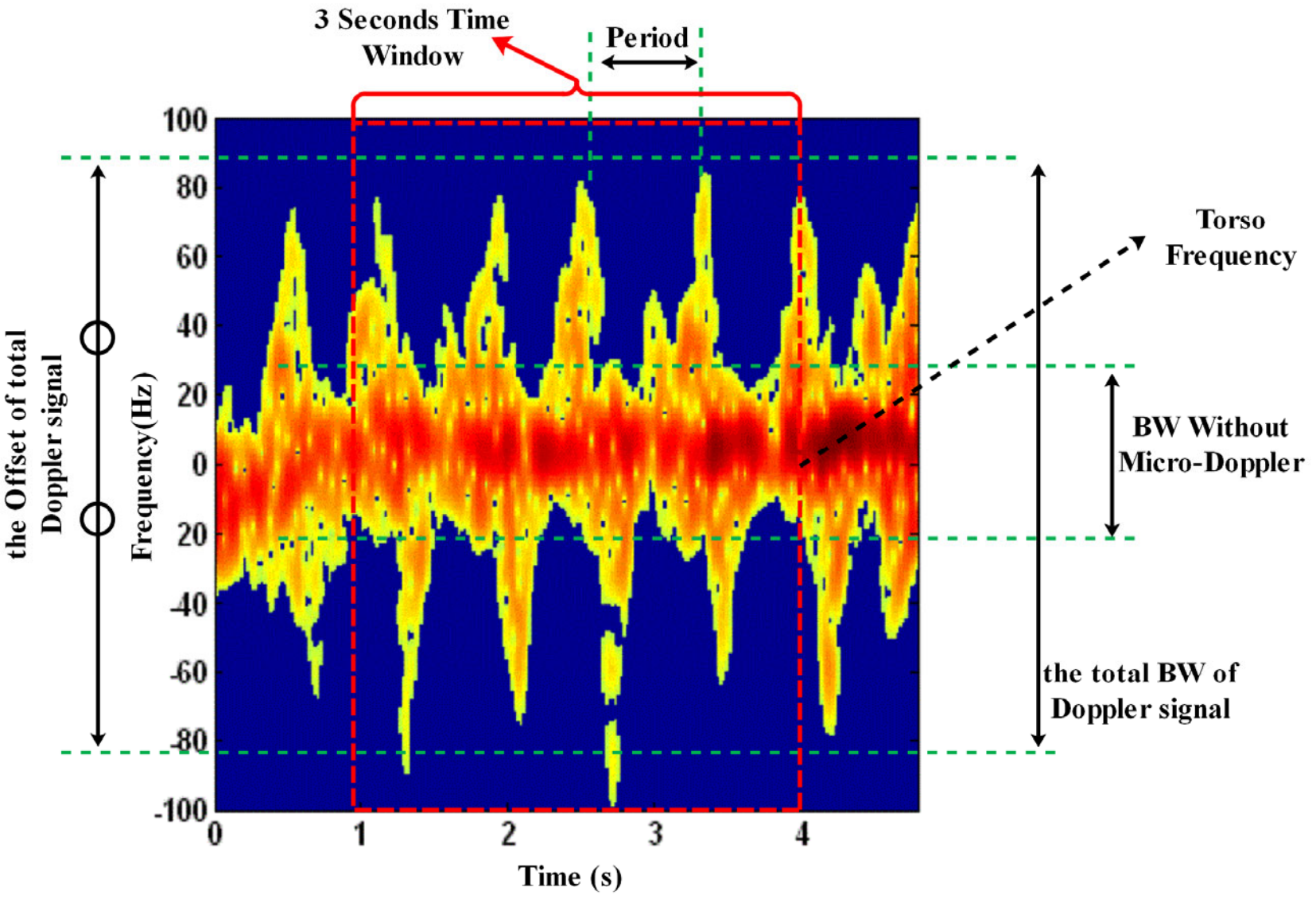

**FIGURE 3. Features illustration of the time-frequency spectrum [24].**

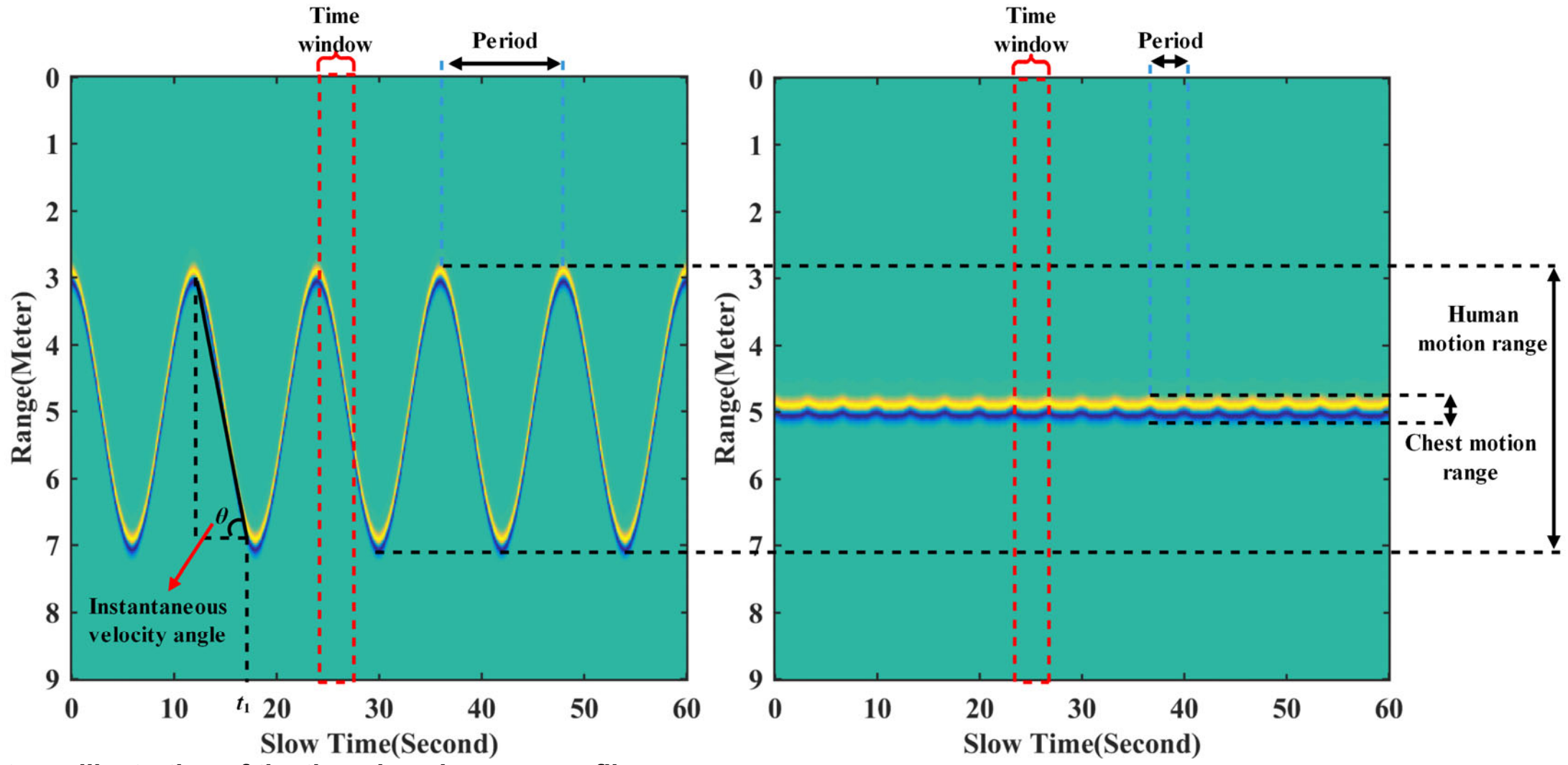

**FIGURE 4. Features illustration of the time-domain range profile.**

To realize the penetration performance of non-metallic obstacles, IR-UWB through-wall radar has a low center frequency and is limited in spatial resolution and micro-Doppler sensitivity, and it has a weak ability to acquire human body movement information, so it can only perform rough recognition of human motion status. Therefore, in this paper, human motion statuses are defined as follows:

Human walking: human moves back and forth in radial and tangential direction within the power of radar detected beam;

Human standing still: human stands still within the power of the radar detected beam. Except for the fretting caused by normal breathing and heartbeat, the human is accompanied by some small -amplitude oscillation.

Considering that the human motion status occurs in the detection scene, then the empty scene (i.e. the human target does not exist in the detection scene) is defined as the human motion status.

## III. RADAR SYSTEM AND DATA ACQUISITION

### A. IR-UWB THROUGH-WALL RADAR SYSTEM

IR-UWB through-wall radar used in this paper is independently designed by the School of Aeronautics and Astronautics, Central South University, China. The structure of the radar is shown in FIGURE 5, and the specific design method of its submodule can be referred to these papers [25], [26]. The improved butterfly antennas are used in this radar system and it can be operated wirelessly with a tablet terminal or with a cable connected to a personal computer. TABLE I shows the key parameters of the IR-UWB radar. The pulse signal generating module generates a trigger pulse with a pulse repetition frequency of 400 KHz, and the center frequency of the transmitted signal is 500 MHz. After the transmitted signal is scatter by the target, the return signals are sampled by the echo acquisition module and processed by the field programmable gate array (FPGA) for the further step. Due to the narrow pulse width of the transmitted signal, it is necessary to use an analog-to-digital converter (ADC) with a high real-time sampling rate for sampling but it may lead to a high cost. Therefore, we use the equivalent sampling technology in the design of the echo acquisition module. The real-time sampling frequency of the ADC is 400 KSa/s and using this method, the sampling frequency can reach 5GSa/s. The sample numbers of the received signal are 768 in the fast time direction. In the slow time direction, 8 return signals are received in 1 second. To improve the signal noise-ratio (SNR) of signal, the average processing of multiple sampling at a single point is carried out. Therefore, the record time of a single frame echo signal can be calculated as follows:

$$t_s = \frac{1}{400kHz} \times M_s \times M + \sum_{m=0}^{M-1} m \times 200(ps) \times M_s \qquad (4)$$

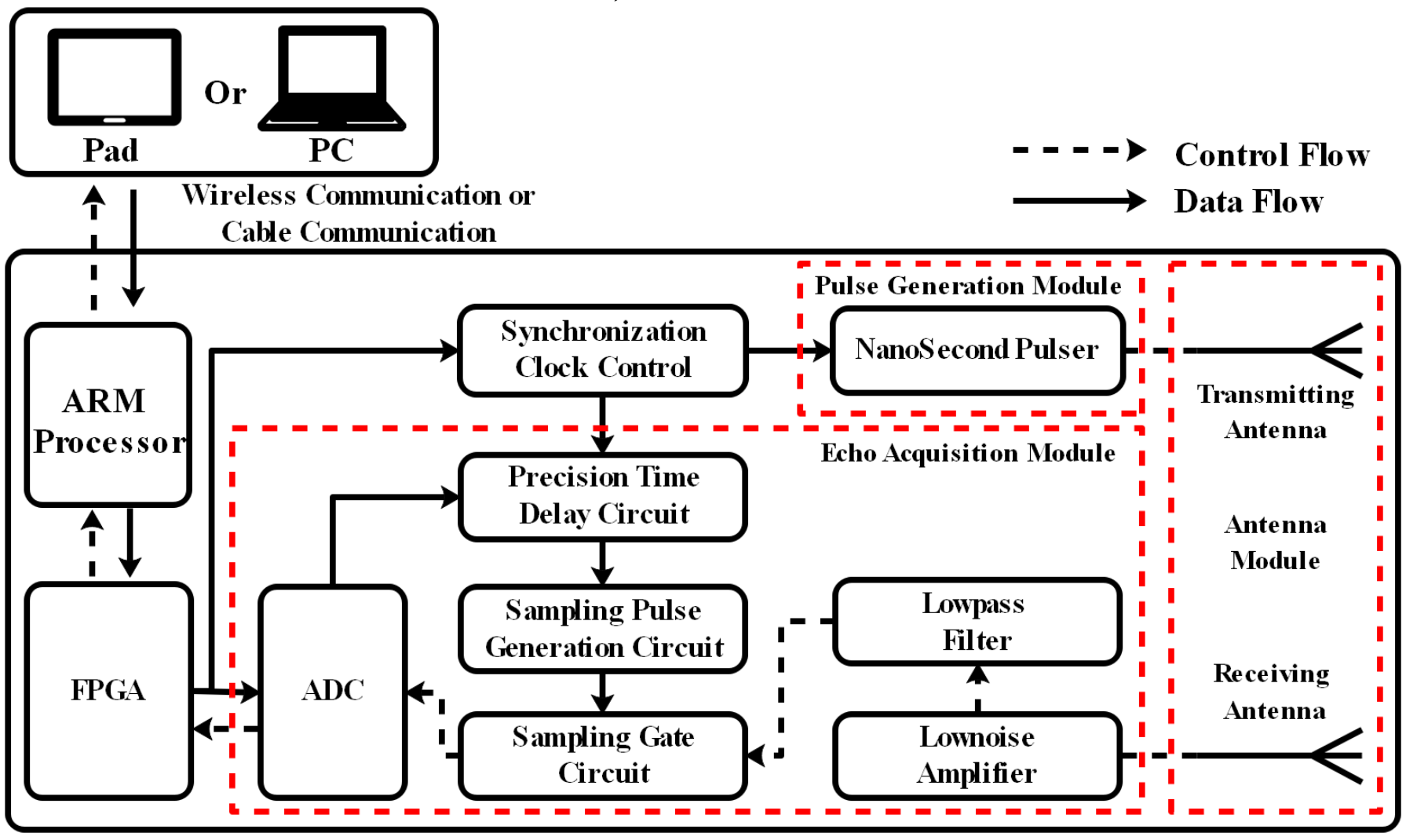

**FIGURE 5.** **The structure of IR-UWB through wall radar**

TABLE I
THE PARAMETERS OF IR-UWB THROUGH-WALL RADAR

| Parameter | Value |
|---|---|
| The distance of transceiver antennas | 0.15 meters |
| Operating mode | Impulse radio |
| Central frequency | 500MHz |
| Pulse width | 2 Nanosecond |
| Sampling points of signal | M=768 |
| Sampling interval by equivalent sampling | 200ps |
| Sampling numbers in single point | Ms=64 |
| Pulse repetition frequency | 400KHz |
| ADC real-time sampling rate | 400KSa/s |

### *B. TIME-DOMAIN RANGE PROFILE DATA ACQUISITION*

The acquisition of time-domain range profile data is obtained by the actual measurement of the human motion statuses using IR-UWB through-wall radar, and the corresponding experimental scenes are shown in FIGURE 6. The IR-UWB radar is close to the wall, and the height of radar is 1.5 meters from the ground. The dielectric material penetrated is a brick-concrete wall with a thickness of 0.2 meters. In the experimental setup, the human walking range is 3 to 8 meters. The volunteers participating in the experiment were 3 healthy adults. Based on the above experimental conditions, the dataset contains a total of 4230 time-domain range profile data with pixel 768×32 and the observation time of each sample is 4 seconds.

TABLE II and TABLE III are the time-domain range profile data of different humans at different times and in different motion statuses. By analyzing the time-domain range profile data sample from these tables, the following conclusions can be obtained:

1) When the human in different motion status, the corresponding time-domain range profile is different, which is roughly the same as the time-domain range profile obtained by simulation;

2) When the human in the same motion status, the corresponding time-domain range profile features are consistent. In other words, in the same motion status, the time-domain range profile features are invariant and independent of a human target;

3) When penetrating different media, the time-domain range profile is still invariant. Because the different media mainly causes the distortion of electromagnetic wave propagation speed, amplitude, and other parameters, and finally affects the accurate positioning of the human;

4) The longer the observation time is, the more sufficient the description of the human motion status is, which is more conducive to the recognition of the human motion status, but it will affect the recognition efficiency.

In conclusion, the construction of the IR-UWB through-wall radar human motion status dataset is completed. FIGURE 7 shows the composition of the human motion status dataset. The construction of the data set provides data support for the realization of subsequent human motion status recognition algorithm. In the following section in this paper, we introduce a human motion status recognition algorithm.

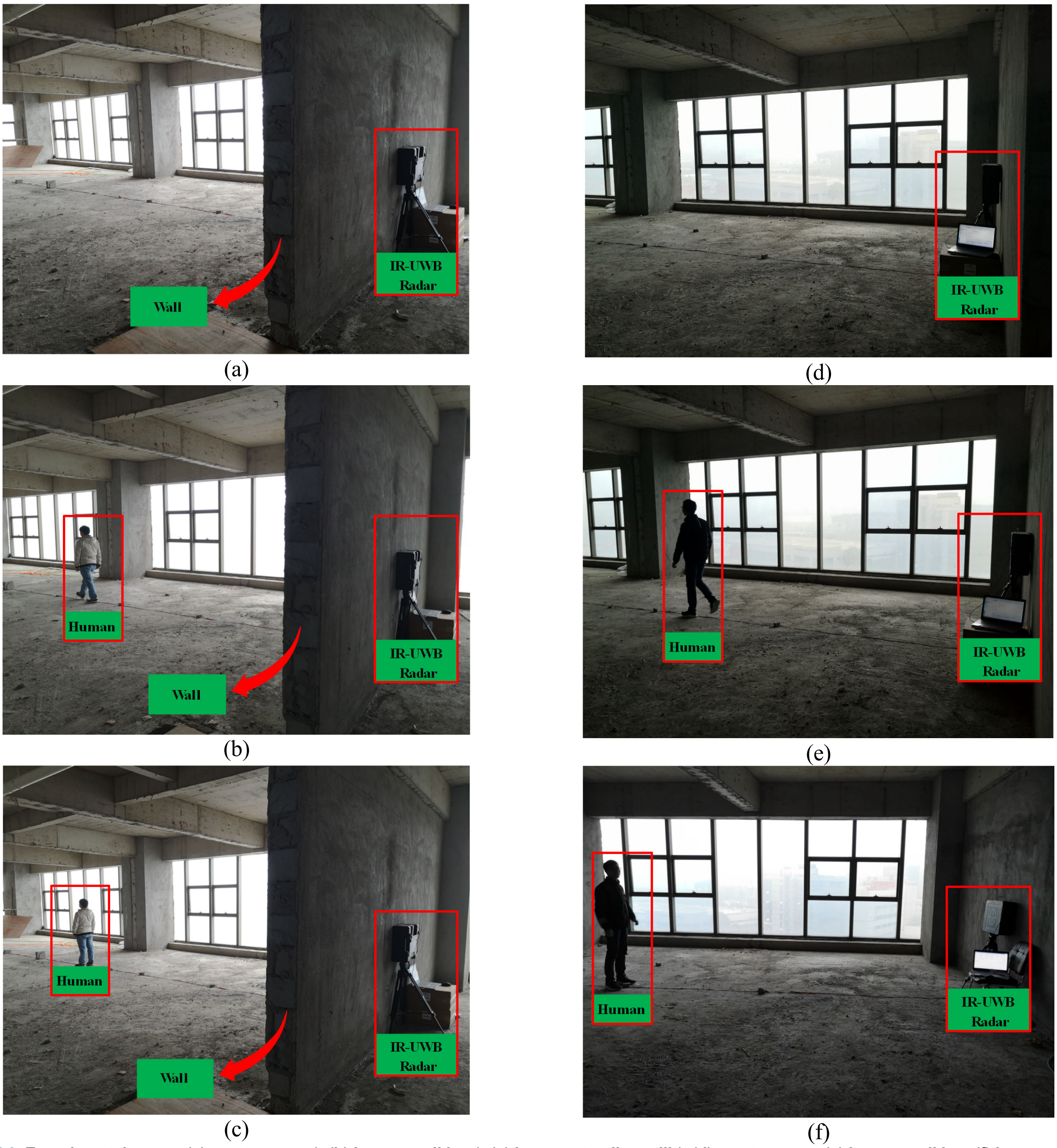

**FIGURE 6.** **Experimental scene. (a) empty scene *. (b) human walking *. (c) human standing still *. (d) empty scene. (e) human walking. (f) human standing still. (* denote the experiment is carried out under the condition of through wall).**

## IV. A Demo: Human Motion Status Recognition Using Convolutional Neural Networks (CNN) for IR-UWB Through-Wall Radar

Based on the dataset collected in Section 3, In this section, we introduce how to use the dataset to recognize the human motion status. This section mainly provides a dataset division ration from the perspective of machine learning. Based on the divided dataset, CNN is trained to realize the recognition of human motion status.

### *A. DATASET DIVISION*

From the constructed dataset, it needs to be divided. Generally, the dataset is divided into a training set and a test set. The training set is used to train the model to learn the task and adjust the parameters of the model, the test set is used to evaluate the performance of the trained model. In short, the training set enables the model to abstract general rules, while the test set is used to check whether the rules abstracted from the model are correct. Generally speaking, the following suggestions should be considered in the division of dataset:

1) The selection of the test set and the training set should be mutually exclusive as far as possible [27], and the sample of the test set should not be used in the training set. If the sample of the test set is used as the sample of the training set, the evaluation result of the whole model will be too optimistic.

2) The division of the test set and training set should keep the proportion of data categories similar. Taking this dataset constructed in Section 3 as an example, when dividing the training set, the number of samples corresponding to each human motion status should be equal as possible. If the sample imbalance occurs in the training set, the model will tend to the class with more samples, which is not conducive to the generalization of the model.

3) The number of samples in the training set and data set should maintain a certain proportion, this suggestion is considered from the overall dataset. Reference [27] mentioned that when the dataset is divided into a training set and test set, it will face the dilemma of deviation-variance. Therefore, $\frac{2}{3} - \frac{4}{5}$ of the dataset is used as the training set, and the rest data is used as the test set.

4) The division should consider the training set and test set samples from the same distribution. Besides, the training set also needs to be divided into a part of the validation set to adjust the supper parameters of the trained model.

To sum up, the IR-UWB human motion status dataset is divided as follows: the ratio of the training set, validation set, and test set is 6:2:2, as shown in **FIGURE 8**, the corresponding sample numbers are 2592, 864 and 864.

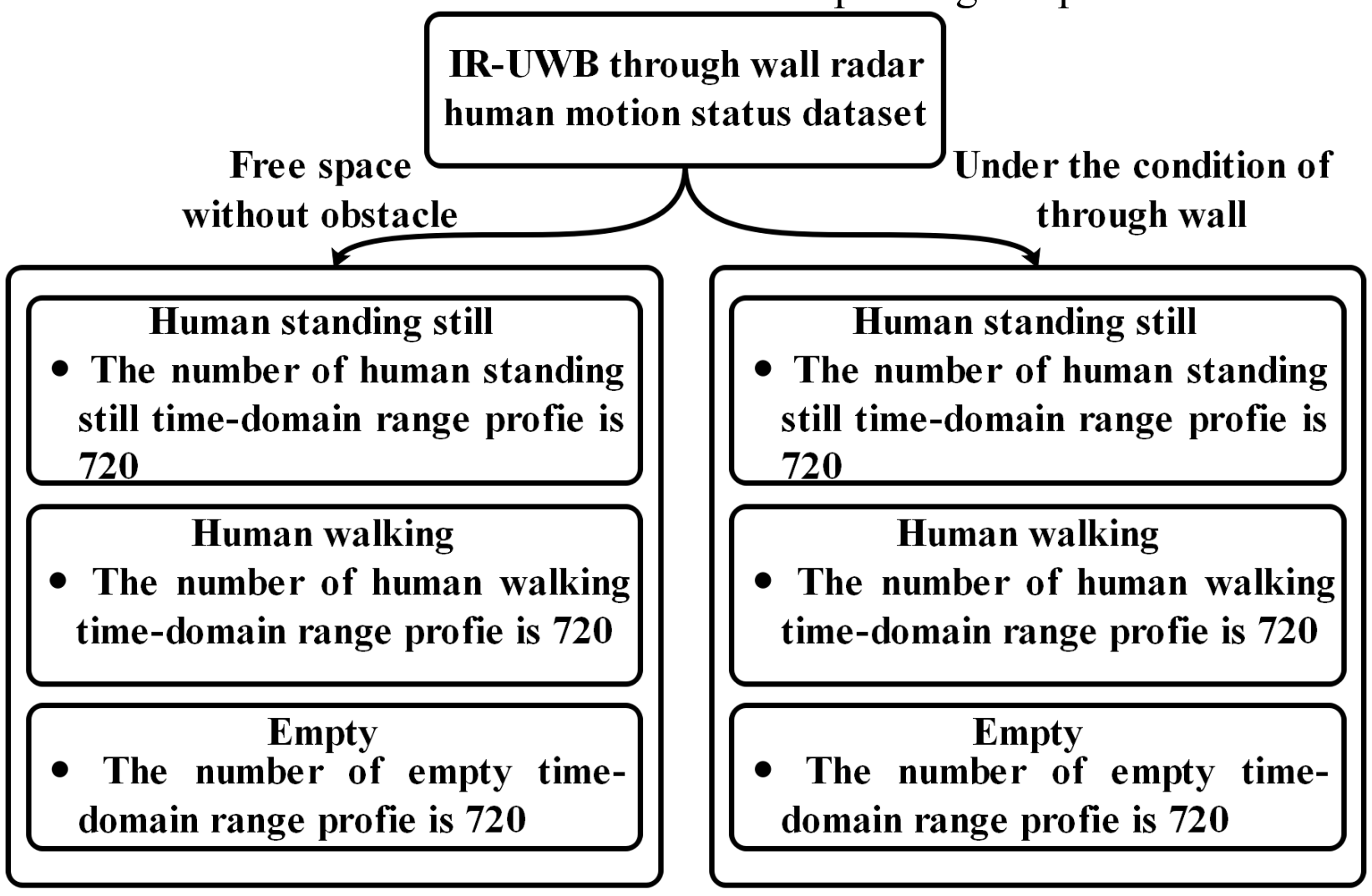

**FIGURE 7.** The composition of the IR-UWB through wall radar human motion status dataset.

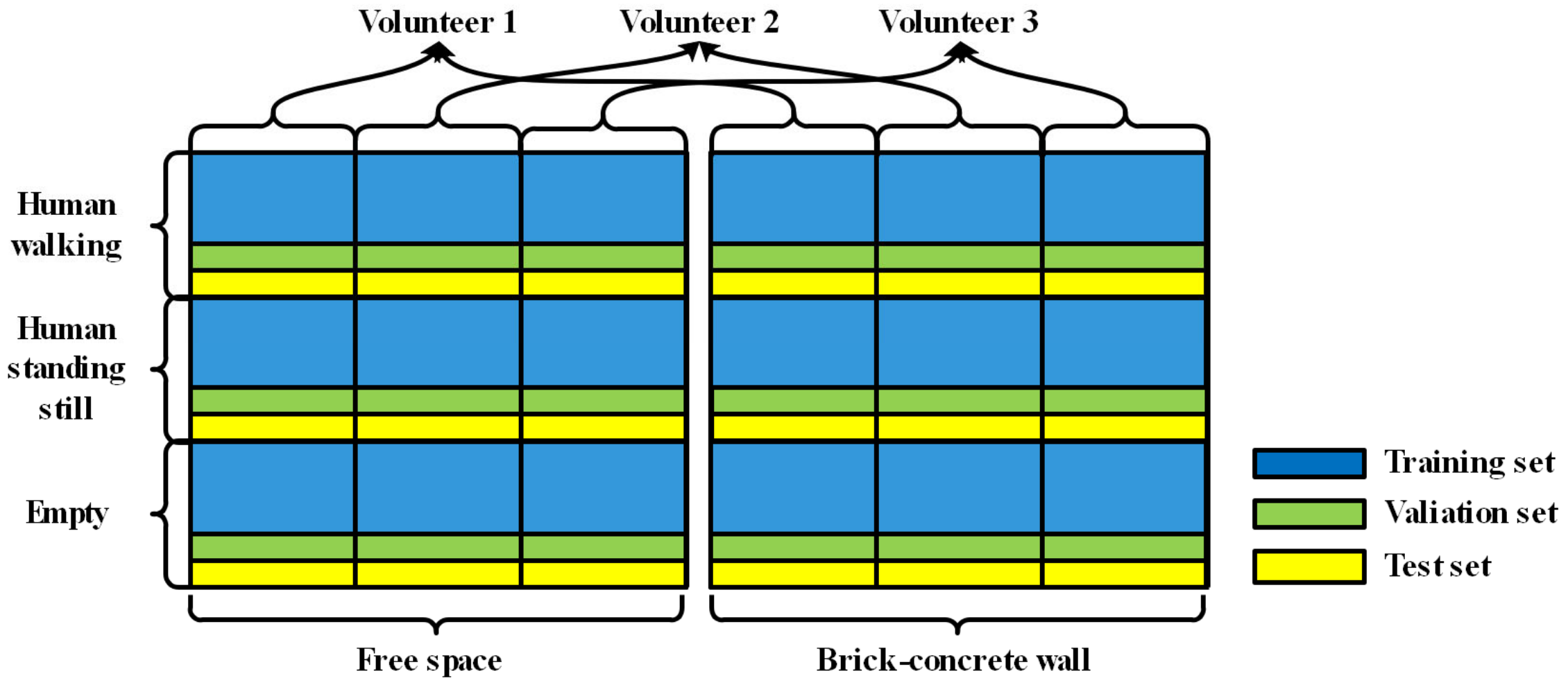

**FIGURE 8.** The division of IR-UWB through wall radar human motion status dataset.

TABLE II
TIME-DOMAIN RANGE PROFILES OF VOLUNTEER NO.1 IN DIFFERENT MOTIONS UNDER DIFFERENT PENETRATING MEDIUMS

| Volunteer ID | Penetrating medium | Human walking | Human standing still | Empty |
|---|---|---|---|---|
| Volunteer NO.1 | Free space | 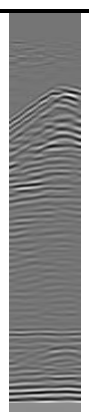 | 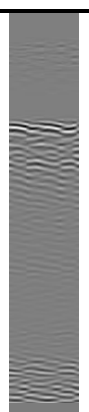 | 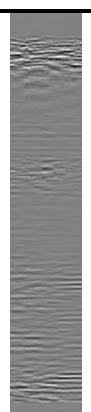 |
| | Brick-concrete wall | 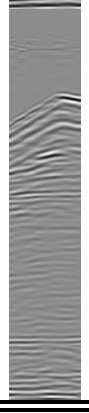 | 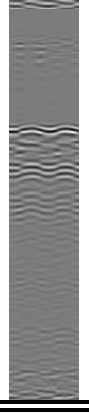 | 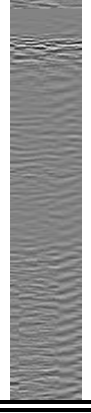 |

TABLE III
TIME-DOMAIN RANGE PROFILES OF VOLUNTEER NO.2 IN DIFFERENT MOTIONS UNDER DIFFERENT PENETRATING MEDIUMS

| Volunteer ID | Penetrating medium | Human walking | Human standing still | Empty |
|---|---|---|---|---|
| Volunteer NO.2 | Free space | 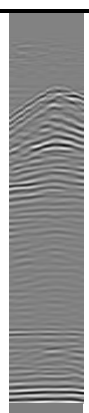 | 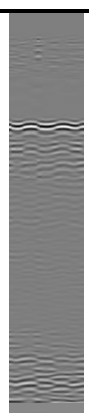 | 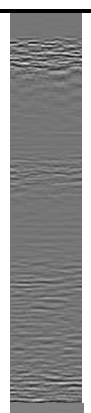 |
| | Brick-concrete wall | 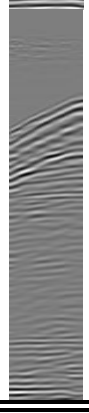 | 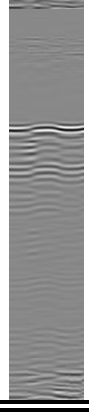 | 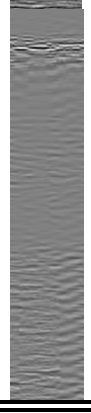 |

### *B. HUMAN MOTION STATUS RECOGNITION BASED ON CNN*

#### 1) CONVOLUTIONAL NEURAL NETWORKS (CNN) THEORY AND DESIGN

Convolutional neural networks (CNN) [28], [29] is one of the deep learning algorithms widely used in computer vision. Driven by large data, the recognition or classification results are directly output via end-to-end deep CNN. FIGURE 9 is the CNN used to target recognition by doppler radar [30]. The essence of CNN is a hierarchical model and its infrastructure is the combination of a convolutional layer, pooling layer and full connected layer. By adding the convolutional layer, and pooling layer, the network depth is deepened and more local features are obtained. This framework is based on the conclusion of Hubel's research on the neurons of the cat's visual cortex [31]: in the process of human visual brain cognition, low-level specific features are continuously transferred to high-level abstract features via neurons. CNN uses multiple convolutional layers and pooling layers to simulate the human visual brain cognitive mode and obtain local features. Then the local features are integrated into the form of a full connected layer. In the convolutional layer, the local features are obtained by the inner product operation of the convolutional kernel and input data, and the nonlinear transformation of the activation function. Here are some common activation functions: Sigmoid function, Than function, Relu function. Convolutional kernels with different sizes can be interpreted as feature extraction with different scales for input data. The zero-padding operation is often performed to ensure the size consistency of input data. The main purpose of the pooling layer is to reduce the dimension of feature data and restrain the overfitting of the network. The pooling layer reduces the number of parameters of the trained model and improves the operational efficiency of the network model. The common pooling methods are Max pooling, Average pooling, and LP-norm pooling. In the full connected layer, the softmax function is used to solve the multi-classification problem. Assuming that there are $k$ classification problem, the output of the softmax function can be calculated as follows:

$$Output = \begin{bmatrix} P(y=1|\ x;W_1,b_1) \\ P(y=2|\ x;W_2,b_2) \\ \cdots \\ P(y=k|\ x;W_k,b_k) \end{bmatrix} = \frac{1}{\sum_{j=1}^{k} e^{(W_j x+b_j)}} \begin{bmatrix} e^{(W_1 x+b_1)} \\ e^{(W_2 x+b_2)} \\ \cdots \\ e^{(W_k x+b_k)} \end{bmatrix} \quad (5)$$

where the $W=[W_1,W_2,...W_k]$ and $b=[b_1,b_2,...b_k]$ are the corresponding weights and biases. The training process of CNN includes two parts: forward propagation and error back-propagation. The training goal is to minimize the loss function of the whole network. The error back-propagation updating model parameters driven by error. The methods like SGD, Momentum, Adam, RmsProp are proposed to improve the update speed of parameters.

In this paper, we designed the CNN for IR-UWB through-wall radar human motion status recognition. TABLE IV is the specific parameter setup of CNN. The design inspiration of the CNN comes from the structure form of VGGNet [32], namely the combination of multiple convolutional layers and pooling layer, and the number of the convolutional kernel in each layer shows a certain change rules, as shown in FIGURE 10.

TABLE IV
THE PARAMETERS SETUP OF CNN

| Layer | Parameters | Value |
|---|---|---|
| Input layer | Input size | 768×32 |
| Convolutional layer (C1) | Kernel size | 3×3 |
| | Filter number | 16 |
| | Activation function | Relu |
| Convolutional layer (C2) | Kernel size | 3×3 |
| | Filter number | 32 |
| | Activation function | Relu |
| Convolutional layer (C3) | Kernel size | 3×3 |
| | Filter number | 64 |
| | Activation function | Relu |
| Convolutional layer (C4) | Kernel size | 2×2 |
| | Filter number | 128 |
| | Activation function | Relu |
| Pooling layer (P5) | Pooling size | 2×2 |
| | Step | 2 |
| Full connected layer (F6) | Number of neurons | 128 |
| | Activation function | Relu |
| Full connected layer (F7) | Number of neurons | 64 |
| | Activation function | Relu |
| Output layer | Number of neurons | 3 |
| | Classifier | Softmax |

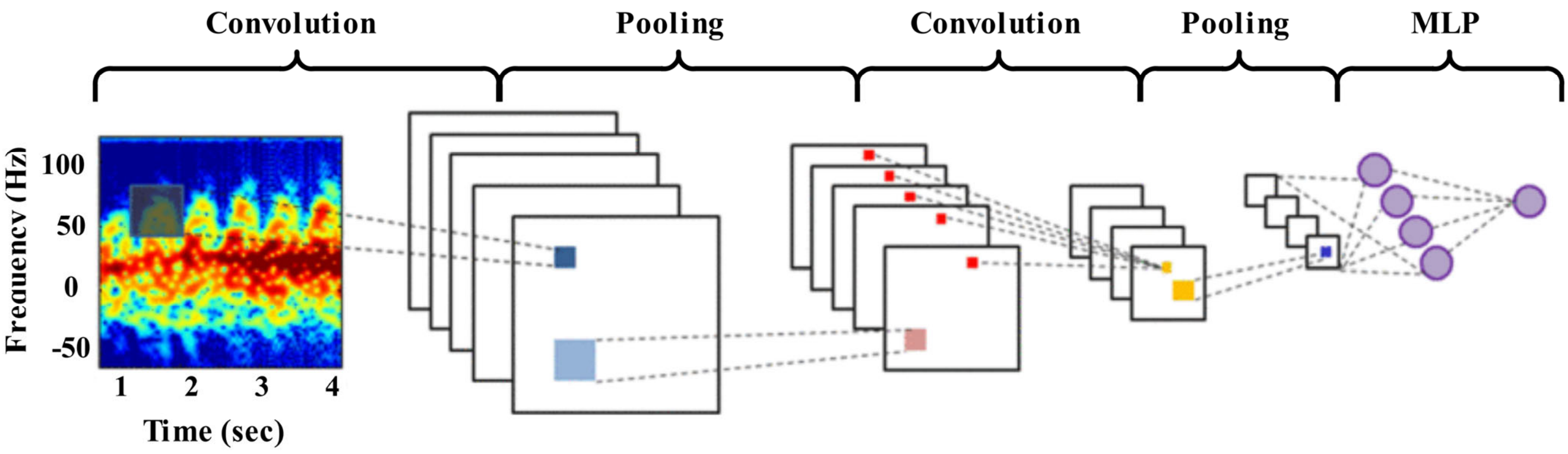

FIGURE 9. The CNN used to doppler radar target recognition proposed in [30].

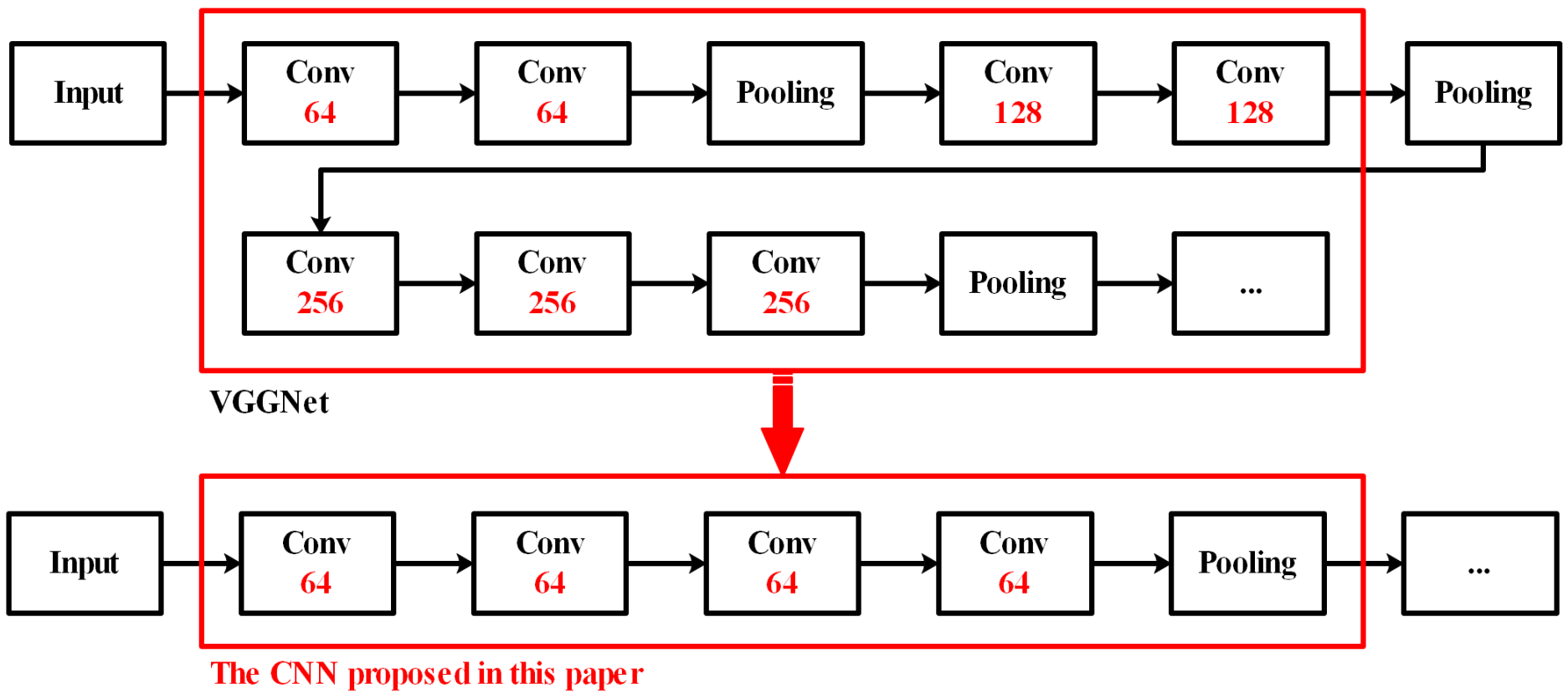

FIGURE 10. The structure comparison between the VGGNet and CNN which proposed in this paper.

### 2) THE PROCESS OF HUMAN TARGET MOTION STATUS RECOGNITION AND RESULTS ANALYSIS

According to the description of the data set divided in Section IV.A and the principle of CNN in Section IV.B.1, the flowchart of human motion status recognition based on CNN is as follows:

Step. 1: read the training set samples of the data set, get the time-domain range profile and its corresponding label, and normalize the time-domain range profile globally;

Step. 2: Using the time-domain range profile as the input sample of the CNN. Set up the parameters of CNN such as activation function type, iteration times, network structure. In the training process, CNN automatically extracts the features of the time-domain range profile. Finally, the output of the network output layer is used to calculate the error between the predicted output and the actual output, and the error back-propagation to update the parameters. The weight and bias of the network are updated by using the accelerated gradient descent algorithm (RMSprop). This is repeated until the cost function meets the iterative conditions, and the training is completed to get the appropriate target operation;

Step. 3: fixed and saved the trained CNN model of human motion statue recognition;

Step. 4: read the test set samples of the data set, processed them as described in step. 1, and input them into the well trained CNN in step. 3. Then recognition of the human motion state can be obtained.

FIGURE 11 is the flow chart of the target motion status recognition algorithm based on CNN. The CNN was implemented with Keras running on top of Tensorflow. The CNN was performed on the computer with 3.6 GHz Intel(R) Core (TM) i9-9900K CPU (64G RAM) and GeForce RTX 2080Ti GPU. In the training process of recognition model, the training was performed for 50 epochs with a batch size of 100, the learning rate is 0.001. We calculated and recorded categorical cross-entropy as the loss function when the training stage. FIGURE 12 shows the recognition accuracy and categorical cross-entropy value of the training set. It can be seen from the figure that with the increase of iteration times, the accuracy of the training set gradually tends to be stable and the cross-entropy loss tends to converge.

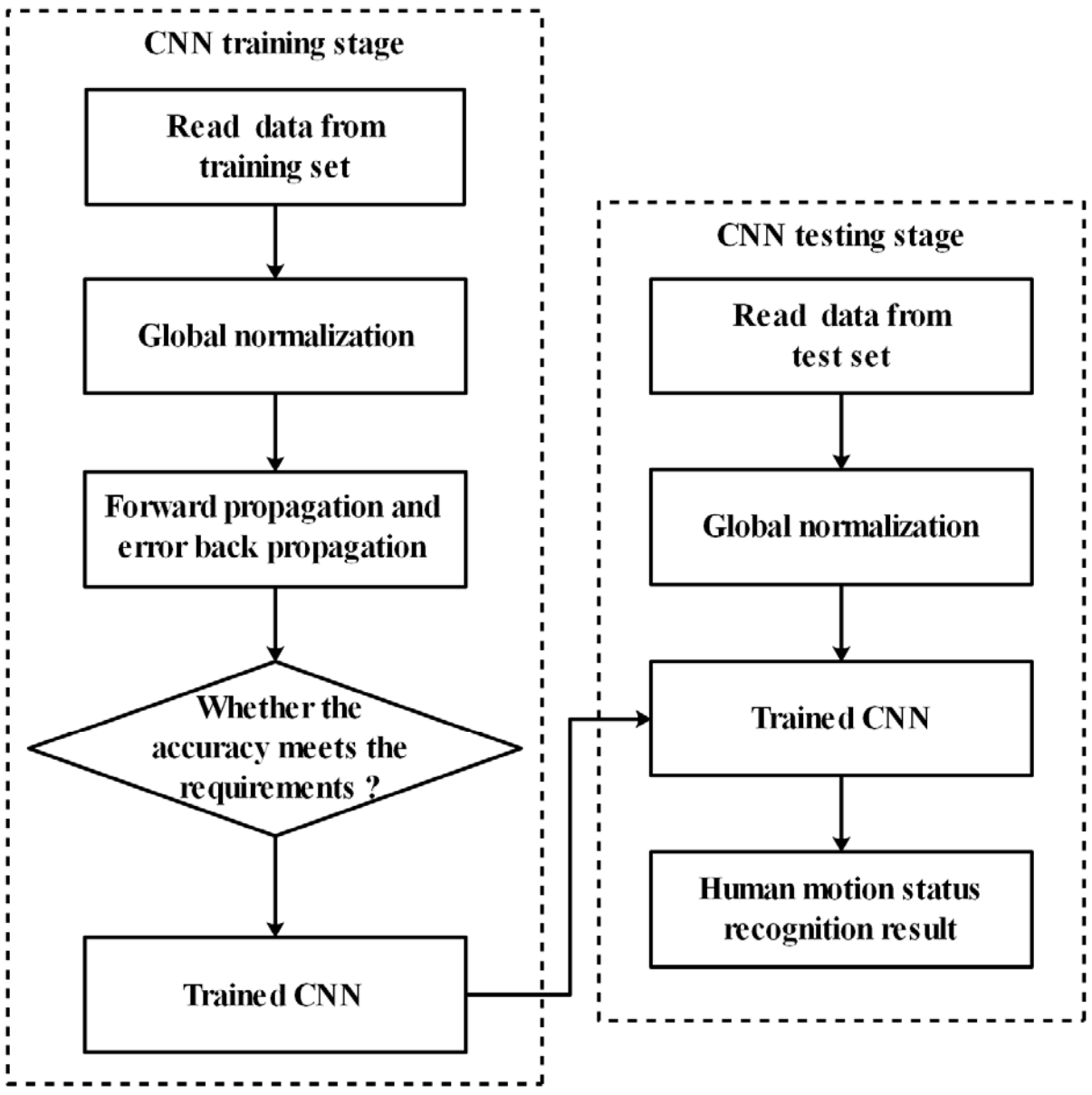

**FIGURE 11.** The flow chart of human motion status recognition based on CNN.

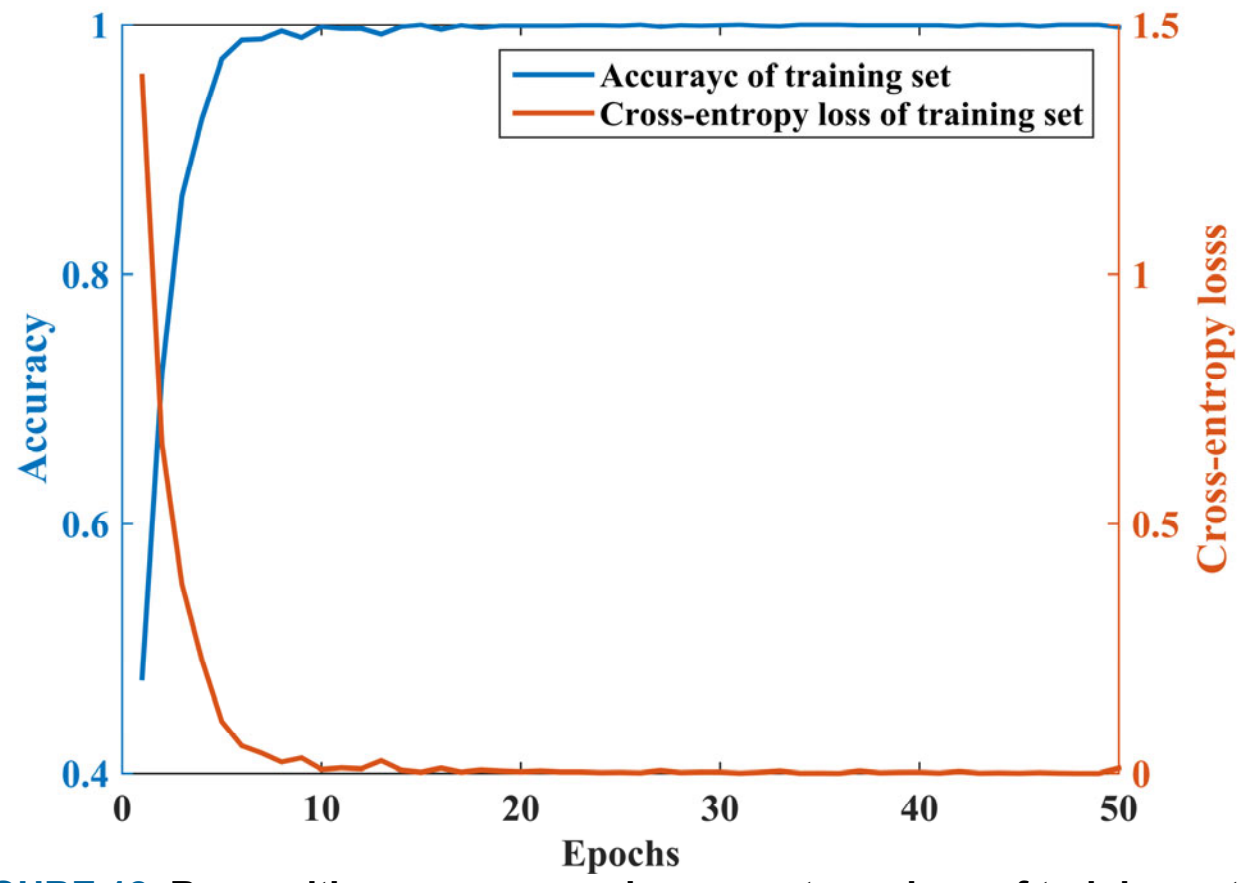

**FIGURE 12.** Recognition accuracy and cross-entropy loss of training set.

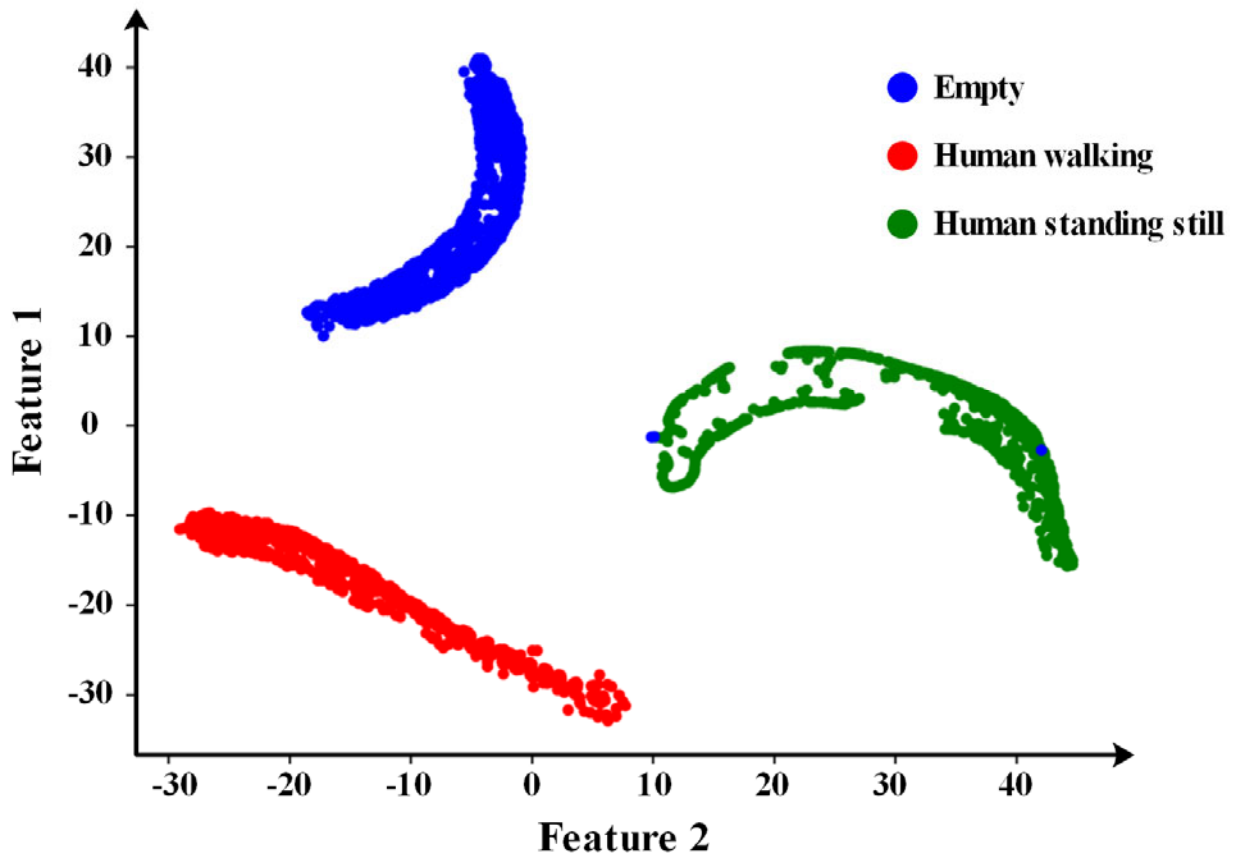

**FIGURE 13.** Features of t-SNE mapping method.

To verify the feature extraction ability of the training network, the t-distributed stochastic neighbor embedding (t-SNE) algorithm [33] is used to map the high-dimensional features extracted by trained CNN set to two-dimensional for feature visualization. As shown in FIGURE 13, it can be seen that the trained CNN can well extract the features of three kinds of motion status of human targets. TABLE V shows the performance results of the trained CNN in the test set. The first row in the table represents the actual motion state of the sample, and row 2 to 4 represents the predicted state of the network. The recognition accuracy of the training model is 99.7%. Especially, 100% accuracy can be achieved for human walking and empty scene.

TABLE V
THE CONFUSITON MATRIX OF THE TEST SET DATA

| Predict motion | Human walking# | Human standing still# | Empty# | Sum |
|---|---|---|---|---|
| Human walking | 288 | 1 | 0 | 288 |
| Human standing still | 0 | 287 | 0 | 288 |
| Empty | 0 | 0 | 288 | 288 |
| Accuracy | 100% | 99.7% | 100% | |

# means these are actual labels.

## V. CONCLUSIONS AND FUTURE PROSPECTS

In this paper, the dataset of human motion status based on the IR-UWB through-wall radar is published, hoping to promote the further integration of the UWB through-wall radar human target recognition and machine learning algorithm. Based on the human motion model of UWB radar, the time-domain range profile simulation of humans in different motion statuses is performed. Using the simulation results, the time-domain range profile features under different motion status are analyzed and to guide the human target motion state in the actual experimental situation, namely human walking, human standing still and empty. For the published datasets, this paper proposes a method of human motion status based on CNN. By using the trained CNN, the recognition accuracy of human motion status can reach 99%. Our results show that machine learning has potential in human target recognition of through-wall radar. Although the provided cases have a good recognition performance, we need to realize that the method based on CNN may have high computational complexity. Therefore, in practical application deployment, we should focus on the complexity and real-time of the CNN model.

In future research, we think that the work of human target recognition of UWB through-wall radar should be deeply explored from the following directions:

1) In the dataset proposed in this paper, due to the definition of human motion status is relatively simple, the feature description of time-domain range profile is adopted, and this method depends on the time length. However, if the selected time length is too long, two motion states may overlap in the same observation window and reduce the recognition efficiency. Therefore, it necessary to explore the method of human target motion status recognition based on the instantaneous features, which can fundamentally remove the dependence on time length and can estimate the duration of motion status. This kind of method is real-time and will be helpful to further study the fine recognition of human motion state.

2) At present, most of UWB through-wall radar work in low frequency and are limited in spatial resolution and Doppler sensitivity to ensure their detection ability in non-metallic and non-transparent media attenuation, so it is difficult to extract the spatial information of human limbs. Therefore, it is necessary to improve the ability of human information acquisition at low frequency. Combined with the human motion scattering model [34], [35], the robust feature representation method for describing fine human motion state should be further studied. Meanwhile, we also need to focus on the multi-dimensional feature extraction related to humans and study the identification algorithm of human target based on the human motion state recognition.

3) The research of the UWB through-wall radar system and its algorithm are stated in the 1980s and has made great progress. However, there is a lack of the UWB through-wall radar dataset for other scholars to study, and the recognition algorithm based on machine learning often needs sufficient data to complete the algorithm development. Therefore, we call on relevant organizations to cooperate in UWB radar data acquisition in complex scenes, to promote the further development of UWB through-wall radar human target recognition.

## APPENDIX

The dataset mentioned in this article is available online at the GitHub in https://github.com/ZhengliangZhu-2020/IR-UWB-Through-wall-Radar-Human-Motion-Status-Dataset.

## ACKNOWLEDGMENT

This work was supported by the National Key Research and Development Program of China under Grant 2018YFC0810202. The authors would like to thank the editor and anonymous reviewers for their valuable ideas, time, and suggestions for improving the quality of the manuscript.

## REFERENCES

[1] E. J. Baranoski, "Through-wall imaging: Historical perspective and future directions," *Journal of the Franklin Institute*, vol. 345, no. 6, pp. 556–569, Sep. 2008, doi: 10.1016/j.jfranklin.2008.01.005.

[2] K. Tan, S. Wu, J. Chen, Z. Xia, F. Guangyou, and S. Meng, "Improved human respiration detection method via ultra-wideband radar in through-wall or other similar conditions," *IET Radar, Sonar & Navigation*, vol. 10, no. 3, pp. 468–476, Mar. 2016, doi: 10.1049/iet-rsn.2015.0159.

[3] F. H. C. Tivive, A. Bouzerdoum, and M. G. Amin, "A Subspace Projection Approach for Wall Clutter Mitigation in Through-the-Wall Radar Imaging," *IEEE Trans. Geosci. Remote Sensing*, vol. 53, no. 4, pp. 2108–2122, Apr. 2015, doi: 10.1109/TGRS.2014.2355211.

[4] M. Le and B. Van Nguyen, "Multivariate Correlation of Higher Harmonics for Heart Rate Remote Measurement Using UWB Impulse Radar," *IEEE Sensors J.*, vol. 20, no. 4, pp. 1859–1866, Feb. 2020, doi: 10.1109/JSEN.2019.2950635.

[5] D. Yang, Z. Zhu, and B. Liang, "Vital Sign Signal Extraction Method Based on Permutation Entropy and EEMD Algorithm for Ultra-Wideband Radar," *IEEE Access*, vol. 7, pp. 178879–178890, 2019, doi: 10.1109/ACCESS.2019.2958600.

[6] S. Z. Gurbuz and M. G. Amin, "Radar-Based Human-Motion Recognition With Deep Learning: Promising Applications for Indoor Monitoring," *IEEE Signal Process. Mag.*, vol. 36, no. 4, pp. 16–28, Jul. 2019, doi: 10.1109/MSP.2018.2890128.

[7] Ma *et al.*, "An Accurate Method to Distinguish Between Stationary Human and Dog targets Under

Through-Wall Condition Using UWB Radar," *Remote Sensing*, vol. 11, no. 21, p. 2571, Nov. 2019, doi: 10.3390/rs11212571.

[8] Y. Ma *et al.*, "Multiscale Residual Attention Network for Distinguishing Stationary Humans and Common Animals Under Through-Wall Condition Using Ultra-Wideband Radar," *IEEE Access*, vol. 8, pp. 121572–121583, 2020, doi: 10.1109/ACCESS.2020.3006834.

[9] L. Jiang, X. Zhou, L. Che, S. Rong, and H. Wen, "Feature Extraction and Reconstruction by Using 2D-VMD Based on Carrier-Free UWB Radar Application in Human Motion Recognition," *Sensors*, vol. 19, no. 9, p. 1962, Apr. 2019, doi: 10.3390/s19091962.

[10] L. Jiang, G. Wei, and L. Che, "Human Motion Recognition Method By Radar Based on CNN," *Computer Applications and Software*, vol. 36, no. 11, 2019.

[11] L. Jiang, C. Li, and L. Che, "Human Motion Recognition by-Ultra-wide Band Radar," *JOURNAL OF ELECTRONIC MEASUREMENT AND INSTRUMENTATION*, vol. 32, no. 1, pp. 129–133, 2018.

[12] J. D. Bryan, J. Kwon, N. Lee, and Y. Kim, "Application of ultra-wide band radar for classification of human activities," *IET Radar Sonar Navig.*, vol. 6, no. 3, p. 172, 2012, doi: 10.1049/iet-rsn.2011.0101.

[13] F. Qi, F. Liang, H. Lv, C. Li, F. Chen, and J. Wang, "Detection and Classification of Finer-Grained Human Activities Based on Stepped-Frequency Continuous-Wave Through-Wall Radar," *Sensors*, vol. 16, no. 6, p. 885, Jun. 2016, doi: 10.3390/s16060885.

[14] F. Qi, H. Lv, F. Liang, Z. Li, X. Yu, and J. Wang, "MHHT-Based Method for Analysis of Micro-Doppler Signatures for Human Finer-Grained Activity Using Through-Wall SFCW Radar," *Remote Sensing*, vol. 9, no. 3, p. 260, Mar. 2017, doi: 10.3390/rs9030260.

[15] H. Du, T. Jin, Y. He, Y. Song, and Y. Dai, "Segmented convolutional gated recurrent neural networks for human activity recognition in ultra-wideband radar," *Neurocomputing*, vol. 396, pp. 451–464, Jul. 2020, doi: 10.1016/j.neucom.2018.11.109.

[16] J. Deng, W. Dong, R. Socher, L.-J. Li, Kai Li, and Li Fei-Fei, "ImageNet: A large-scale hierarchical image database," in *2009 IEEE Conference on Computer Vision and Pattern Recognition*, Miami, FL, Jun. 2009, pp. 248–255, doi: 10.1109/CVPR.2009.5206848.

[17] H. Du, T. Jin, Y. Song, and Y. Dai, "Unsupervised Adversarial Domain Adaptation for Micro-Doppler Based Human Activity Classification," *IEEE Geosci. Remote Sensing Lett.*, vol. 17, no. 1, pp. 62–66, Jan. 2020, doi: 10.1109/LGRS.2019.2917301.

[18] Y. Lang, Q. Wang, Y. Yang, C. Hou, D. Huang, and W. Xiang, "Unsupervised Domain Adaptation for Micro-Doppler Human Motion Classification via Feature Fusion," *IEEE Geosci. Remote Sensing Lett.*, vol. 16, no. 3, pp. 392–396, Mar. 2019, doi: 10.1109/LGRS.2018.2873776.

[19] W. Wang, M. Zhang, and L. Zhang, "Classification of Data Stream in Sensor Network With Small Samples," *IEEE Internet Things J.*, vol. 6, no. 4, pp. 6018–6025, Aug. 2019, doi: 10.1109/JIOT.2018.2867649.

[20] Y. Li, W. Wang, and Y. Jiang, "Through Wall Human Detection Under Small Samples Based on Deep Learning Algorithm," *IEEE Access*, vol. 6, pp. 65837–65844, 2018, doi: 10.1109/ACCESS.2018.2877730.

[21] "CMU Graphics Lab Motion Capture Database." http://mocap.cs.cmu.edu/.

[22]Y. Fei, *Ultra Wideband Radar Theory and Technology*. Beijing: Nation Defense Industry Press, 2010.

[23] Xin Li, Dengyu Qiao, Ye Li, and Huhe Dai, "A novel through-wall respiration detection algorithm using UWB radar," in *2013 35th Annual International Conference of the IEEE Engineering in Medicine and Biology Society (EMBC)*, Osaka, Jul. 2013, pp. 1013–1016, doi: 10.1109/EMBC.2013.6609675.

[24] Youngwook Kim and Hao Ling, "Human Activity Classification Based on Micro-Doppler Signatures Using a Support Vector Machine," *IEEE Trans. Geosci. Remote Sensing*, vol. 47, no. 5, pp. 1328–1337, May 2009, doi: 10.1109/TGRS.2009.2012849.

[25] F. Zhang and B. Liang, "UWB radar life detector system design and test," *Fire Science and Technology*, vol. 35, no. 7, pp. 967–969, 2016.

[26] Z. Zhu, B. Liang, D. Yang, and R. Zhao, "Research, fabrication and application of IR-UWB through wall radar system," *Transducer and Microsystem Technologies*, vol. 39, no. 4, pp. 103–106, 2020.

[27] Z. Zhou, *Maching Learning*, Tsinghua University Press. Beijing, 2016.

[28] Y. Lecun, "Gradient-Based Learning Applied to Document Recognition," *Proceedings of the IEEE*, vol. 86, no. 11, pp. 2278–2323, 1998.

[29] Y. LeCun, B. Boser, and J. Denker, "Backpropagation Applied to Handwritten Zip Code Recognition," *Neural Computation*, vol. 1, no. 4, pp. 541–551, Dec. 1989, doi: 10.1162/neco.1989.1.4.541.

[30] Y. Kim and T. Moon, "Human Detection and Activity Classification Based on Micro-Doppler Signatures Using Deep Convolutional Neural Networks," *IEEE GEOSCIENCE AND REMOTE SENSING LETTERS*, vol. 13, no. 1, p. 5, 2016, doi: 10.1109/LGRS.2015.2491329.

[31] D. H. Hubel and T. N. Wiesel, "Receptive fields, binocular interaction and functional architecture in the cat's visual cortex," *The Journal of Physiology*, vol. 160, no. 1, pp. 106–154, Jan. 1962, doi: 10.1113/jphysiol.1962.sp006837.

[32] K. Simonyan and A. Zisserman, "Very Deep Convolutional Networks for Large-Scale Image Recognition," presented at the International Conference on Learning Representations, 2015.

[33] L. van der Maaten and G. Hinton, "Visualizing Data using t-SNE," *Journal of Machine Learning Research*, 2008.

[34] Y. Zhang, Y. Zhu, Y. Cheng, and X. Li, “Human Target Radar Echo Signal Analysis Based on Micro-Doppler Characteristic,” *Signal Processing*, vol. 25, no. 10, pp. 1616–1623, Oct. 2009.
[35] F. He, “Researches on key techniques of wide band/ultra-wide band radar detection and signature extraction of locomotor human targets,” Ph.D dissertation, National University of Defense Technology, Changsha, Hunan, China, 2011.